# Evidence of abrupt changes in the orbital periods of two cataclysmic variables

David Boyd

**Abstract**

We report evidence from eclipse timing that two otherwise apparently normal cataclysmic variables have experienced abrupt changes in their orbital periods. The orbital period of HS 2325+8205 increased by 22.15 msec ($1.3 \times 10^{-6}$ of the orbital period) around 1 February 2011 and the orbital period of EP Dra increased by 4.51 msec ($7.2 \times 10^{-7}$ of the orbital period) around 7 December 2001. In neither case was there any apparent change in the subsequent behaviour of the system. These changes are currently unexplained.

**Introduction**

Cataclysmic variables (CVs) are binary systems in which a white dwarf (WD) accretes hydrogen-rich material from a main sequence (MS) companion via Roche lobe overflow through the inner Lagrangian point (L1). The accreting stream of material forms a disc around the WD before eventually settling onto its surface. The gradual accumulation of material in the disc may eventually lead to thermal instability which causes the disc to brighten for a period before fading back to its quiescent state. These bright events are known as outbursts and are a signature of the sub-class of CVs known as dwarf novae (DNe). If the WD has a magnetic field, formation of a disc may be partially inhibited or, if the magnetic field is sufficiently strong, prevented altogether in which case accretion flows directly onto one or both of the magnetic poles of the WD. These are known as AM Her stars or polars.

**HS 2325+8205**

First reported as variable by Morgenroth (1936), HS 2325+8205 (hereafter HS2325) was subsequently identified as a possible CV in the Hamburg Quasar Survey (Aungwerojwit et al. 2006). In quiescence its magnitude is around 17 rising to slightly brighter than 14 in outburst. It experiences eclipses which are 0.4 magnitude deep during outburst and around one magnitude deep in quiescence. The presence of eclipses indicates a relatively high orbital inclination. Its light curve shows no evidence of standstills between outburst and quiescence so it is unlikely that HS2325 is a UGZ type of DN. Outbursts all appear to have very similar maximum brightness with no evidence of brighter superoutbursts so it is unlikely to be a UGSU type DN. We are therefore drawn to the conclusion that HS2325 is a U Gem or UGSS type DN (Hellier 2001). HS2325 was the subject of a paper by Pyrzas et al. (2012) which found an inclination of 75° and an orbital period of 4.6640263(3) hr. This puts HS2325 above the period gap and therefore likely to be experiencing magnetic braking.

**Observations**

Unfiltered photometry of HS2325 from the 1.2m telescope at the Kryoneri Observatory in Greece between 2004 and 2006 was obtained from Prof Boris Gänsicke. These data were phased on the above orbital period and 9 eclipses were identified. The first Kryoneri eclipse was used as T_zero for the ephemeris and given cycle number zero. To find the time of minimum for these and subsequent eclipses the lower half of each eclipse was fitted with a quadratic polynomial. Uncertainty on the time of minimum was derived from the uncertainties in the photometry measurements.

A total of 4526 V-band and unfiltered magnitude measurements of HS2325 by various observers between 2007 and 2021 were obtained from the AAVSO International Database (AAVSO AID). A list of observers who contributed more than 10 measurements is given in Table 1. Measurements were obtained using a variety of telescopes up to 0.4m aperture. These data were searched for time series which recorded eclipses and could be analysed to provide a time of minimum. In total 18 such eclipses were found, all between 2007 and 2009. All but one of these were recorded by the late Ian Miller.

A further 6208 V-band and unfiltered magnitude measurements were made by the author between 2009 and 2022 with a 0.35m telescope. Magnitudes were calculated by differential aperture photometry with respect to the magnitudes of comparison stars from the AAVSO chart for HS2325. These provided a further 62 eclipse timings which are listed in Table 2. Subsequently, three eclipse timings between January 2012 and August 2013 were found in Hardy et al. (2017). This made a total of 92 eclipses with measured timings. The times of all eclipses were converted to Heliocentric Julian Date (HJD).

Online databases of All-Sky Automated Survey for Supernovae (ASAS-SN) (Kochanek et al. 2017), Zwicky Transient Facility (ZTF) (Masci et al. 2019) and Asteroid Terrestrial-impact Last Alert System (ATLAS) (Tonry et al. 2018) were searched for information on eclipses but the cadence of these measurements meant they were not able to provide any eclipse timings.

**O-C analysis**

Knowing the orbital period, cycle numbers could be unambiguously assigned to each eclipse and a linear fit performed of eclipse time vs cycle number. This provides a calculated time for each eclipse. Observed minus calculated (O-C) times for the 35 eclipses between 2004 and 2011 are consistent with the following linear ephemeris with E as the cycle number. The number in parentheses indicates the uncertainty in the last digit.

$$T_{min} (HJD) = 2453168.46156(5) + 0.194334482(8) * E \qquad (1)$$

The 54 eclipse timings recorded by the author from 2012 onwards were clearly divergent from this ephemeris. These, together with the 3 eclipses from Hardy et al. in 2012 and 2013, defined a new linear ephemeris with a longer orbital period.

$$T_{min} (HJD) = 2453168.45836(11) + 0.194334739(4) * E \qquad (2)$$

Figure 1 shows the O-C plot for all 92 eclipses with respect to the linear ephemeris of equation (1). Both linear ephemerides are shown as dotted lines with published eclipse timings in black and new timings by the author in red. The intersection of the linear ephemerides occurred around 1 February 2011 (JD 2455594) at which point the orbital period of HS 2325 appears to have increased abruptly by 22.15 msec, a fractional change of $1.3 \times 10^{-6}$ of the orbital period.

We also considered the possibility that the change may have occurred gradually and tried to fit a quadratic polynomial to the data to model this. As expected, this gave a much poorer fit to the data than two straight lines.

To see if this apparently major change in the dynamics of the binary had any effect on its subsequent behaviour, we compare in Figures 2 and 3 magnitude measurements of HS2325 excluding eclipses from before the change (HJD 2454833 to 2454963) and after the change (HJD 2457479 to 2457841). The quiescent and outburst magnitudes appear to be very similar before and after the change. In the later and more complete set of data we have indicated 25 outbursts, 7 long outbursts with an average duration of 18 days, and 18 short outbursts with an average duration of 13 days. This appears to be entirely normal outbursting UGSS behaviour whose outburst amplitude remained unchanged by the event.

**EP Draconis**

EP Dra was first recognised as an eclipsing CV showing a small amount of circular polarisation at optical wavelengths in Remillard et al. (1991). This classified it as an AM Her binary or polar in which the WD has a sufficiently strong magnetic field that no accretion disc forms and material from the secondary transfers directly along magnetic field lines onto the pole or poles of the WD. It initially had the name H1907+690 from the HEAO 1 survey and exhibited deep eclipses which gave an orbital period of 1.743750(4) hr. EP Dra is therefore below the period gap.

**Observations**

EP Dra is around magnitude 18 out of eclipse and magnitude 20 in eclipse. As eclipses last approximately 7 min, unfiltered photometry exposures were limited to 60 sec with a 0.35m telescope to retain timing resolution. This means the star is only detectable intermittently during eclipse in images and so presents a challenge to measuring the timing of eclipses. The difficulty of measuring individual eclipses was mitigated by recording on average 6 eclipses during each observing season and combining these data to define a mean eclipse profile and hence obtain a consolidated mid-eclipse time for that season. Uncertainty on the derived time of minimum was determined from uncertainties in the photometry measurements.

Published eclipse times were obtained from Remillard et al. (1991), Schlegel & Mukai (1995), Schwope & Mengel (1997) and Bridge et al. (2003). We recomputed mid-eclipse times for Bridge et al. using their mid-ingress and mid-egress times rather than using their values based on phase zero of the Schwope & Mengel ephemeris. We also estimated the mid-eclipse time in Ramsay et al. (2004) for the eclipse they observed in X-rays in October 2002. Observations by the author began in May 2012 and continued until February 2022. Table 3 lists the 11 new seasonal mid-eclipse times measured by the author. Cycle numbers are based on the cycle count established in Remillard et al. As the majority of published eclipse times are in terms of Barycentric Julian Date (BJD), this was adopted for all eclipse timings for EP Dra.

**O-C analysis**

The following linear ephemeris for EP Dra was calculated using published eclipse times and cycle numbers recorded between 1987 and 2002.

$$Tmin (BJD) = 2447681.72989(4) + 0.0726562678(12) * E \qquad (3)$$

It was apparent that the eclipse times measured by the author since 2012 were not consistent with this ephemeris. Ten further eclipse times were discovered in the PhD thesis of Bours (2015) which were consistent with the timings obtained by the author. Together these eclipse times since 2012 defined a new linear ephemeris which represented an increase in the orbital period of 4.51 msec, a fractional change of $7.2 \times 10^{-7}$ of the orbital period.

$$Tmin (BJD) = 2447681.72660(16) + 0.0726563201(13) * E \qquad (4)$$

These linear ephemerides provide a calculated time for each eclipse. Figure 4 shows a plot of observed minus calculated (O-C) times for all eclipses of EP Dra with respect to the linear ephemeris of equation (3). Published eclipse timings are in black and new timings by the author in red. This also shows the two linear ephemerides in equations (3) and (4) as dotted lines which when extrapolated intersect around 7 December 2001 (JD 2452251), the most likely date for the change to have occurred.

The observation of EP Dra by Ramsay et al. in October 2002 occurred some 9 months after the apparent change in orbital period and they did not report anything unusual about the behaviour of the system. Its current behaviour continues to show nothing unusual.

**Speculation on possible causes**

Such an abrupt change in orbital period is inconsistent with mechanisms which have been proposed to explain progressive changes in the orbital periods of CVs such as loss of angular momentum through magnetic braking associated with a magnetized stellar wind (Knigge et al. 2011), or the Applegate mechanism associated with magnetically induced changes in the internal structure of the secondary star (Applegate 1992). To explain such an abrupt change in orbital period requires a more radical change in the components of the binary such as a spontaneous transfer of mass via Roche lobe overflow from the MS star to the WD, or significant mass loss and/or a change in angular momentum.

In the case of HS2325, and assuming reasonable values for the masses of the WD and MS star (0.75 $M_\odot$ and 0.45 $M_\odot$ respectively), the observed change in orbital period could be caused, in the conservative case where total mass and angular momentum are conserved, by the spontaneous transfer of around $5 \times 10^{-7}$ $M_\odot$ from the MS star to the WD (see Hilditch 2001 for relevant equations). Such a transfer would be more than the rate of $\sim 10^{-6}$ $M_\odot$/yr expected at the peak in nova explosions (Ginzburg & Quataert 2021). This mass transfer is also 2 to 3 orders of magnitude greater than the annual mass transfer rate for a CV due to magnetic braking (Ivanova & Taam 2003). If such a large transfer of mass happened in a single event, it would definitely have a noticeable impact on the brightness of the binary system. We did not see any such change and, as we saw in Figures 2 and 3, the behaviour of HS2325 appeared completely unaffected by the event in February 2011.

In the case of EP Dra, using masses of the WD and MS star suggested in Ramsay et al. (0.68 $M_\odot$ and 0.19 $M_\odot$ respectively), the mass transfer involved would be around $6 \times 10^{-8}$ $M_\odot$ in the conservative case and again would have resulted in visible changes which we did not observe.

We therefore conclude that a very rapid transfer of mass is unlikely to have been the cause of the increase in orbital period of either binary.

It is also worth noting that while HS2325 has an accretion disc, EP Dra does not, so an accretion disc is unlikely to be involved in the process. Further, HS2325 is above the period gap and in the magnetic braking regime while EP Dra is below the gap after magnetic braking has ceased to operate.

**Are there other similar cases?**

We searched the O-C diagrams in the PhD thesis of Bours (2015) and the later refereed publication (Bours et al. 2016) which reviewed eclipse times of white dwarfs in 67 close binaries. We did not find any similar examples of abrupt period change between two linear ephemerides which were sustained over several years.

**Summary**

We observed an apparently abrupt increase of 22.15 msec in the 4.66 hour orbital period of the UGSS dwarf nova HS 2325+8205 which occurred around February 2011. We also observed a similarly abrupt increase of 4.15 msec in the 1.74 hour orbital period of the polar EP Dra around December 2001. In neither case was there any apparent change in the subsequent behaviour of the system. We cannot as yet offer any satisfactory explanation for these changes.

**Acknowledgements**

The author is grateful to Prof Boris Gänsicke for providing data from Kryoneri Observatory and to Dr Chris Lloyd for helpful discussions. He wishes to pay tribute to the late Ian Miller whose early observations of HS2325 were crucial in establishing the orbital period before it changed. He also acknowledges with thanks the variable star observations from the AAVSO International Database contributed by observers worldwide and used in this research and the comparison star charts provided by the AAVSO.

| ID | Name |
|---|---|
| BJAA | Boardman, James |
| JSJA | Johnston, Steve |
| MIW | Miller, Ian |
| MNIC | Mishevskiy, Nikolay |
| PVEA | Popov, Velimir |
| SFY | Shears, Jeremy |
| SGOR | Sjöberg, George |
| SRIC | Sabo, Richard |
| SXN | Simonsen, Michael |

Table 1. AAVSO identifiers and names of observers whose AAVSO data of HS2325 were used in this analysis.

| Date | Cycle no | Mid-eclipse time (HJD) | Error (sec) |
|---|---|---|---|
| 5-Apr-09 | 9051 | 2454927.38277 | 25.08 |
| 13-Apr-09 | 9092 | 2454935.35107 | 38.57 |
| 19-Apr-09 | 9123 | 2454941.37518 | 38.72 |
| 26-Apr-09 | 9159 | 2454948.37076 | 31.88 |
| 12-May-11 | 12998 | 2455694.42128 | 13.93 |
| 29-Jun-11 | 13245 | 2455742.42173 | 26.80 |
| 24-Jul-11 | 13374 | 2455767.49086 | 15.11 |
| 9-Aug-11 | 13456 | 2455783.42639 | 11.67 |
| 6-Jun-15 | 20645 | 2457180.49898 | 9.98 |
| 7-Jun-15 | 20650 | 2457181.47053 | 17.18 |
| 2-Oct-15 | 21252 | 2457298.46042 | 22.86 |
| 24-Oct-15 | 21365 | 2457320.41997 | 16.82 |
| 7-Jan-16 | 21751 | 2457395.43364 | 6.94 |
| 14-Jan-16 | 21787 | 2457402.42941 | 9.73 |
| 18-Feb-16 | 21967 | 2457437.40973 | 10.13 |
| 5-Mar-16 | 22050 | 2457453.53940 | 7.43 |
| 31-Mar-16 | 22183 | 2457479.38594 | 9.00 |
| 4-Aug-16 | 22832 | 2457605.50911 | 8.15 |
| 5-Aug-16 | 22837 | 2457606.48080 | 9.14 |
| 25-Nov-16 | 23413 | 2457718.41770 | 13.00 |
| 23-Dec-16 | 23557 | 2457746.40168 | 14.62 |
| 20-Jan-17 | 23701 | 2457774.38601 | 6.53 |
| 13-Feb-17 | 23825 | 2457798.48357 | 14.16 |
| 15-Mar-17 | 23979 | 2457828.41107 | 12.71 |
| 27-Mar-17 | 24041 | 2457840.46015 | 19.25 |
| 9-May-17 | 24262 | 2457883.40790 | 21.85 |
| 21-Sep-17 | 24957 | 2458018.47048 | 4.66 |
| 15-Oct-17 | 25080 | 2458042.37353 | 9.97 |
| 5-Nov-17 | 25188 | 2458063.36195 | 30.35 |
| 24-Nov-17 | 25286 | 2458082.40647 | 10.84 |
| 15-Dec-17 | 25394 | 2458103.39476 | 9.34 |
| 24-Feb-18 | 25759 | 2458174.32687 | 9.69 |
| 6-May-18 | 26125 | 2458245.45351 | 9.02 |
| 2-Jun-18 | 26264 | 2458272.46594 | 28.56 |
| 24-Jun-18 | 26377 | 2458294.42599 | 29.42 |
| 2-Aug-18 | 26578 | 2458333.48698 | 8.10 |
| 3-Aug-18 | 26583 | 2458334.45890 | 13.04 |
| 28-Oct-18 | 27025 | 2458420.35492 | 29.95 |
| 11-Nov-18 | 27097 | 2458434.34668 | 4.38 |
| 14-Jan-19 | 27427 | 2458498.47714 | 10.38 |
| 4-Feb-19 | 27535 | 2458519.46517 | 10.43 |
| 10-Apr-19 | 27869 | 2458584.37307 | 11.89 |
| 26-May-19 | 28106 | 2458630.43032 | 16.72 |
| 20-Jun-19 | 28235 | 2458655.49965 | 19.64 |

| Date | Cycle no | Mid-eclipse time (BJD) | Error (sec) |
|---|---|---|---|
| 25-Jul-19 | 28415 | 2458690.47987 | 8.38 |
| 26-Aug-19 | 28579 | 2458722.35079 | 13.08 |
| 2-Oct-19 | 28770 | 2458759.46871 | 10.84 |
| 16-Oct-19 | 28842 | 2458773.46084 | 11.42 |
| 26-Mar-20 | 29675 | 2458935.34198 | 17.70 |
| 31-Mar-20 | 29701 | 2458940.39455 | 11.78 |
| 6-Apr-20 | 29732 | 2458946.41872 | 12.35 |
| 31-May-20 | 30015 | 2459001.41548 | 20.14 |
| 12-Dec-20 | 31018 | 2459196.33355 | 26.52 |
| 14-Dec-20 | 31029 | 2459198.47109 | 24.11 |
| 15-Jan-21 | 31193 | 2459230.34209 | 18.62 |
| 22-Feb-21 | 31389 | 2459268.43154 | 11.85 |
| 4-Jun-21 | 31914 | 2459370.45741 | 17.73 |
| 7-Oct-21 | 32557 | 2459495.41484 | 20.95 |
| 21-Mar-22 | 33406 | 2459660.40434 | 14.37 |
| 22-Mar-22 | 33411 | 2459661.37643 | 12.43 |
| 24-Mar-22 | 33421 | 2459663.31940 | 12.02 |
| 8-Jul-22 | 33967 | 2459769.42663 | 31.13 |

Table 2: New mid-eclipse times of HS2325 measured by the author.

| Year | Cycle no | Mid-eclipse time (BJD) | Error (sec) |
|---|---|---|---|
| 2012 | 116185 | 2456123.30155 | 21.80 |
| 2013 | 120591 | 2456443.42524 | 30.68 |
| 2014 | 126339 | 2456861.05383 | 20.35 |
| 2015 | 130541 | 2457166.35523 | 25.84 |
| 2016 | 134634 | 2457463.73758 | 33.00 |
| 2017 | 141747 | 2457980.54226 | 37.53 |
| 2018 | 145128 | 2458226.19315 | 20.74 |
| 2019 | 149173 | 2458520.08793 | 36.12 |
| 2020 | 153640 | 2458844.64366 | 20.74 |
| 2021 | 159289 | 2459255.07921 | 36.35 |
| 2022 | 164230 | 2459614.07388 | 20.74 |

Table 3: New annual mid-eclipse times of EP Dra measured by the author.

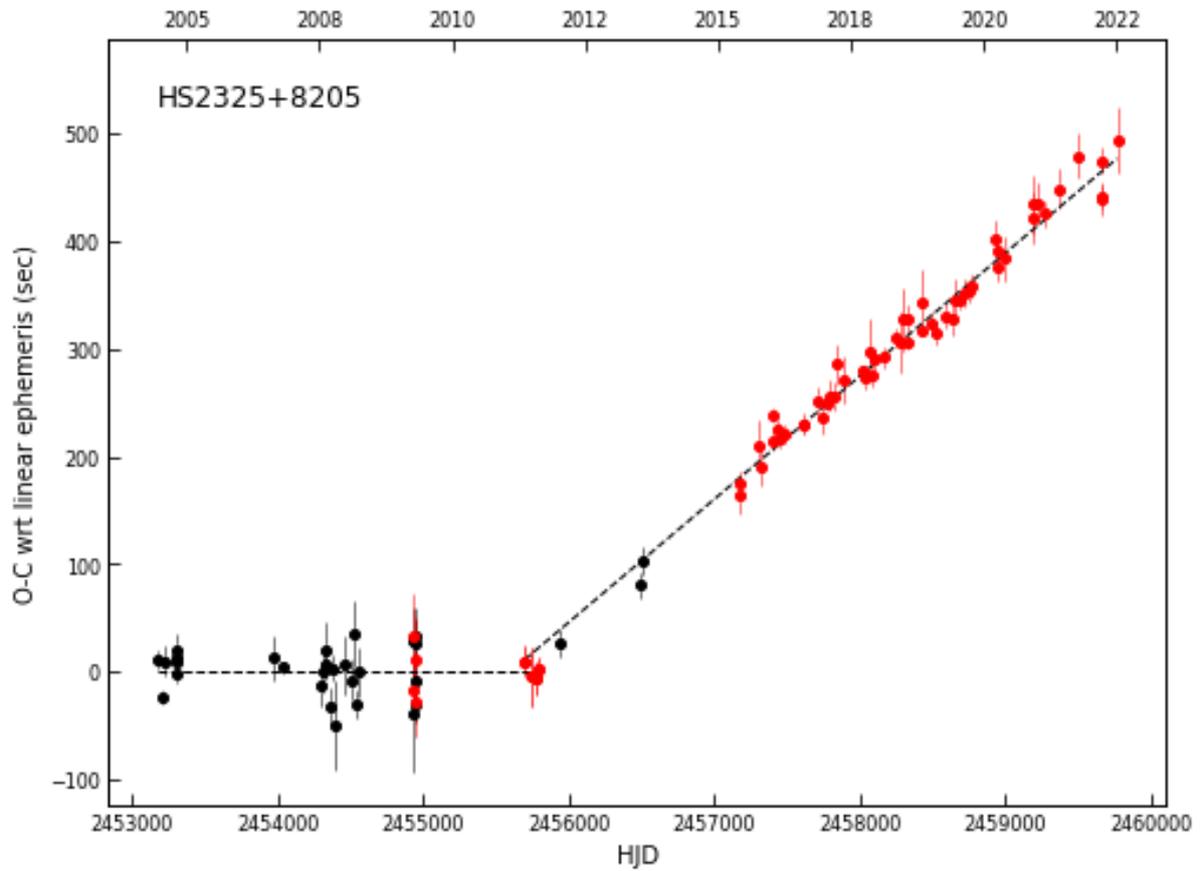

Figure 1. O-C plot for all eclipses of HS2325 with respect to the linear ephemeris in equation (1). Published eclipse timings are in black, those by the author in red.

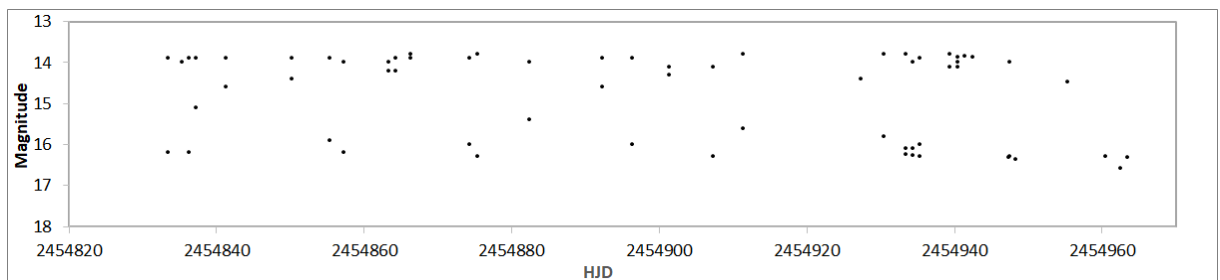

Figure 2: Magnitude measurements of HS2325 excluding eclipses between HJD 2454833 and 2454963 before the change in orbital period.

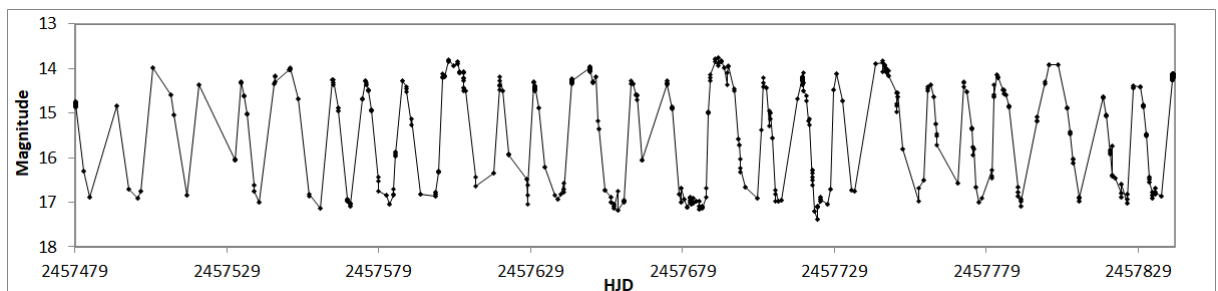

Figure 3. Magnitude measurements of HS2325 excluding eclipses between HJD 2457479 and 2457841 after the change in orbital period showing normal UGSS outbursting behaviour.

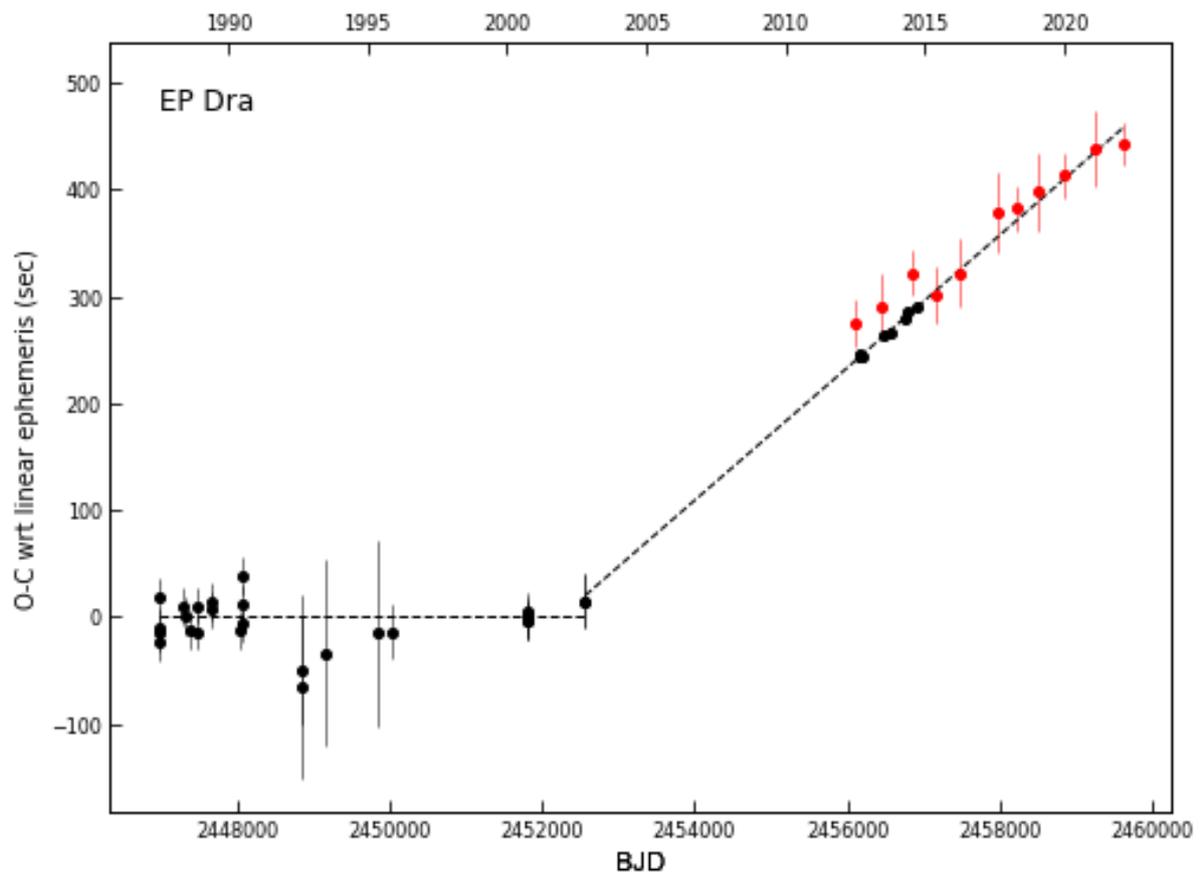

Figure 4. O-C plot for all eclipses of EP Dra with respect to the linear ephemeris in equation (3). Published eclipse timings are in black, those by the author in red.